# THE TECHNOLOGY OF USING A DATA WAREHOUSE TO SUPPORT DECISION-MAKING IN HEALTH CARE

Dr. Osama E.Sheta[1] and Ahmed Nour Eldeen [2]

[1,2]Department of Mathematics (Computer Science) Faculty of Science,
Zagazig University, Zagazig, Elsharkia, Egypt.
[1]oesheta75@gmail.com, [2]ahmednour_cs@yahoo.com

*ABSTRACT:*

*This paper describes the technology of data warehouse in healthcare decision-making and tools for support of these technologies, which is used to cancer diseases. The healthcare executive managers and doctors needs information about and insight into the existing health data, so as to make decision more efficiently without interrupting the daily work of an On-Line Transaction Processing (OLTP) system. This is a complex problem during the healthcare decision-making process. To solve this problem, the building a healthcare data warehouse seems to be efficient. First in this paper we explain the concepts of the data warehouse, On-Line Analysis Processing (OLAP). Changing the data in the data warehouse into a multidimensional data cube is then shown. Finally, an application example is given to illustrate the use of the healthcare data warehouse specific to cancer diseases developed in this study. The executive managers and doctors can view data from more than one perspective with reduced query time, thus making decisions faster and more comprehensive.*

## 1. INTRODUCTION

Managing data in healthcare organizations has become a challenge as a result of healthcare managers having considerable differences in objectives, concerns, priorities and constraints. The planning, management and delivery of healthcare services included the manipulation of large amounts of health data and the corresponding technologies have become increasingly embedded in all aspects of healthcare. Information is one of the most factors to an organization success that executive managers or physicians would need to base their decisions on, during decision making.. Healthcare organizations typically deal with large volumes of data containing valuable information about patients, procedures, treatments and etc. These data are stored in operational databases that are not useful for decision makers or executives.

A majority of database management systems used in these organizations execute online transaction processing (OLTP direct answer to queries at the executive level, such as the what-if and what-next type queries. Decision makers at executive level would like to quickly analyze existing health data and in time to aid in the decision making process. However, stand-alone databases cannot provide such information quickly and efficiently. The concept of data warehousing provides a powerful solution for data integration and information access problems. Data warehousing idea is based on the online analytical processing (OLAP). Basically, this technology supports reorganization, integration and analysis of data that enable users to access information quickly and accurately [1]. The comparison between OLTP and OLAP technology shown in table 1. Finally, OLAP is a tool that used by analyst for planning and decision making. In traditional information systems, businesses have relied on paper-based reports regarding performance in order to make important business decisions. Most of the reports that are created are out dated; these have come as a result of extracting data from operational systems and collating with other sources of data to come up with them. Managers want and need more information, but analysts can provide only minimal information at a high cost within the desired time frames [2]. A healthcare data warehouse designed for this particular purpose is needed.). OLTP databases are designed to process individual records of patients, procedures, treatments, drugs and other similar operations. These databases are updated continually and are suitable to support daily operations. Also these databases cannot provide a

An important role of a healthcare data warehouse is to provide information for Executive manager to analyze situations and make decisions. Put in another way, a healthcare data warehouse provides information for doctors to make decisions and do their jobs more effectively [3]. The top level of decision making involves strategic decision making. At this level managers make decisions about the overall goals of the organization. For instance, types of decisions made on this level include which services need to be provided (such as acute, ambulatory or long term care) and at which geographical location to operate (such as local, state, national). The second level concerns tactical decision making. The decisions made on this level relate to the tactical units of the organization such as patient care services and marketing. The third level concerns the day to day decisions of the organization such as hiring employees, ordering supplies and medications, processing bills [4]. Good systems provide the information needed, so that Managers are making more efficient decisions. Described in this paper is the development and implement of a prototype healthcare data warehouse specific to cancer diseases, employing the new 'data warehouse' technology incorporating large quantity of analysis information needed for healthcare decision-making.

Table 1: Comparison of OLAP and OLTP (adapted from Ref[6])

| Characteristics | OLAP | OLTP |
|---|---|---|
| **Operation** | Analyze | Update |
| **Level of detail** | Aggregate | Detail |
| **Time** | Historical, Current, Projected | Current |
| **Orientation** | Attributes | Records |

## 2. USING A DATA WAREHOUSE

The concept of "data warehousing" arose in mid 1980s with the intention to support huge information analysis and management reporting [5]. Data warehouse was defined According to Bill Inmon a "subject-oriented, integrated, time variant and non-volatile collection of data in support of management's decision making process" [6].

According to Ralph Kimball "a data warehouse is a system that extracts, cleans, conforms, and delivers source data into a dimensional data store and then supports and implements querying and analysis for the purpose of decision making".[7]

Today, data warehouses are not only deployed extensively in banking and finance, consumer goods and retail distribution and demand-based manufacturing, it has also became a hot topic in noncommercial sector, mainly in medical fields, government, military services, education and research community etc.

A data warehouse is typically a read-only dedicated database system created by integrating data from multiple databases and other information sources. A data warehouse is separate from the organization's transactional databases (i.e., OLTP databases). It differs from transaction systems in that [8]:

- It covers a much longer time horizon (several years to decades) than do transaction systems.

- It includes multiple databases that have been processed so that the warehouse's data are subject oriented and defined uniformly (i.e., ''clean prearranged data'').

- It contains non-volatile data (i.e., read-only data) which are updated in planned periodic cycles, not frequently.

- It is optimized for answering complex queries from direct users (decision makers) and applications.

The transactional databases are designed to answer *who* and *what* type questions, they are not very good in answering *what-if, why* and *what-next* type questions. The reason is that data in transactional databases are not necessarily organized to support analytical processing [9].

Data warehouse architecture is a description of the components of the warehouse, with details showing how the components will fit together [10]. Data is imported from several sources and transformed within a staging area before it is integrated and stored in the production data warehouse for further analysis [11]. Since a data warehouse is used for decision making, it is important that the data extracted from multiple sources should be corrected. It is inevitable that when different data are integrated into the data warehouse, there is a high probability of errors and anomalies. Therefore, tools for data extraction, data cleaning, data integration and finally data load are required. Data are stored and managed in the warehouse which presents multidimensional views of data to a variety of front end tools: query tools, report writers, analysis tools, and data mining tools [12]. The architecture of data warehousing is shown in Figure 1:

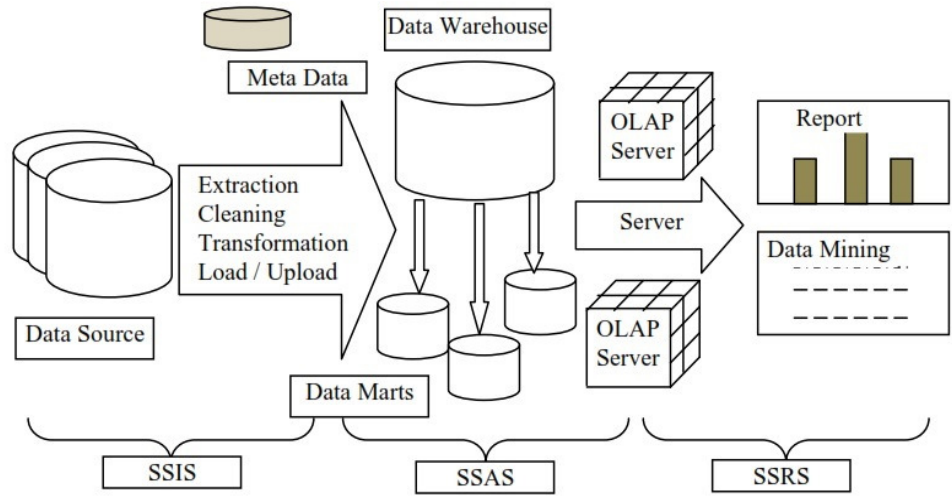

Figure1. Data Warehouse Architecture

SQL Server 2008 is used to develop the Healthcare Data warehouse because it covers not only relational database management service, but also integrated service, analysis service and reporting service. Among them, integration service can help us to integrate data from different heterogeneous data sources, namely providing function of data extraction, transformation and load; analysis services provide us with function of OLAP and data mining which can help us to analyze the current situation and predict the future trend; reporting services provide us with various forms of data report and graphical display of the analysis result.

## 3. DATA WAREHOUSE TECHNOLOGY

### 3.1 Design data warehouse for Cancer diseases

Designing the data warehouse structure is different from designing the operational systems. The operational systems consist of simple pre-defined queries. On the other hand, in data warehousing environments queries join with more tables and more computation time and informality [13]. This leads to an emergence of a new view of data modeling design. As a result of this, the multidimensional or data cube has become the suitable data model for the data warehousing environment. A multidimensional view of the data is important when designing front end tools, database design and query engines for OLAP. Two modeling techniques named "Star Schema" and 'Snowflakes Schema" are used to represent multidimensional data. The 'star' schema is adopted here mainly because of its clarity, convenience and rapid indexing ability [14]. In concise term, a star schema can be defined as a specific type of database design used to support analytical processing, which includes a specific set of denormalized tables. A star schema contains two types of tables show figure 2; this modelling consists of a central table (fact table) and other tables which directly link to it. These tables are known as dimension tables. In general, the fact table contains the keys and measurements. MDX queries then use to query from fact and dimension tables within the star schema, with constraints on the data to return required information.

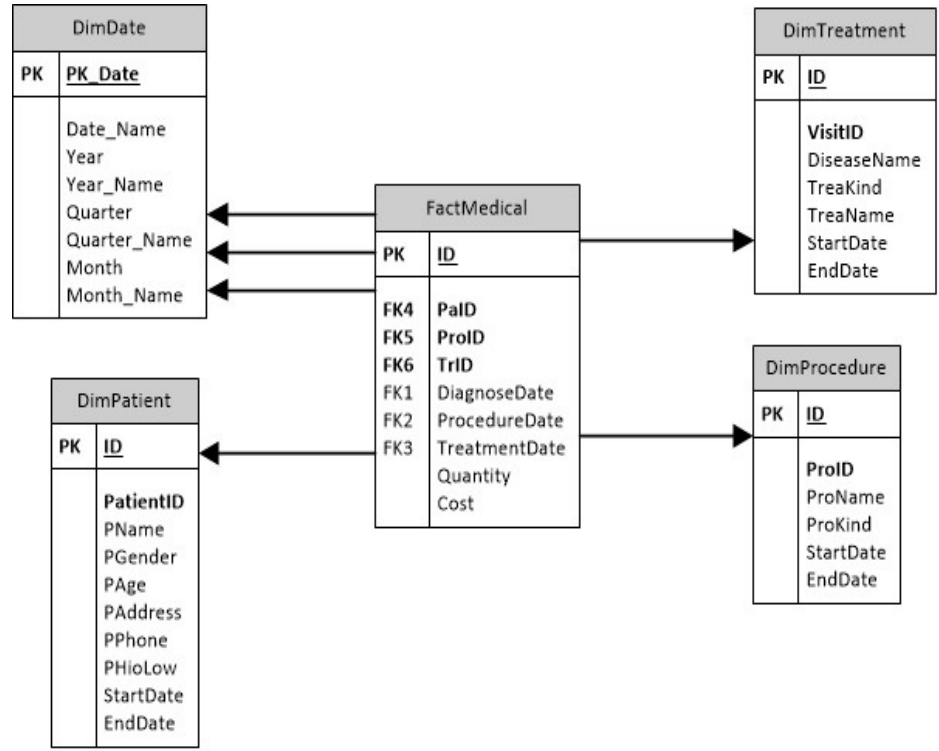

Figure2. Cancer Star Schema

Below describes the fact and dimension tables in detail.

**Fact table:**

The Fact table that describes the subject matter is named FactMedical. The table consists Foreign Keys (PaID, ProID, TrID, DiagnoseDate, ProcedurDate, TreatmentDate) that related with dimensional table and Measures (Cost and Quantity)

**Dimension tables**

The four dimension tables that detailed each entity in the FactMedical table.

Table 2: Dimension tables

| Table name | Table description |
|---|---|
| DimPatient | A table that stores patient information, such as patient name, gender, age, address, phone etc. The data is used to show demographic data for the Cancer disease. |
| DimProcedure | A table that stores all the Procedure such as urine test results, x-ray results and etc and the types of the Procedure |
| DimTreatment | A table that stores all the treatment kind such as Chemotherapy, Radiotherapy and etc. treatment name and disease name. |
| DimDate | A table that stores all date |

## 3.2 Cancer data warehouse development

Cancer data warehouse is developed according to the previous star schema data model. This development contains the following steps:

**Data is extracted from the multi sources, databases and files**
In this step we extracted data from access database which are created in Arabic language from multiple tables such as "Patient, Procedure, Ticket and Visit" tables for transform and load in dimension and fact table**.** Show figure 3 and figure 4.

**Data is transformed and cleaned before loaded.**
Before being loaded into the data warehouse, data extracted from the multiple sources was transformed using built-in transformations contained in SSIS. Transformation means that map extracted data to the same format data for cancer data warehouse repository. We transform Arabic column name to English name and then use the data conversion transformation tool that enables us to convert columns from one data type to another. In the dimension load tasks, the fuzzy look up is used to lookup data rows with spelling mistakes and correct such mistakes. Show figure 5.

**Data is loaded into the data warehouse.**
Whereas the data flows loading the warehouse dimensions are first passed to a sort editor that removes any duplicated rows of data before being passed on to a slowly changing dimension adapter, flows that populate the fact tables are not first passed through a sort adapter before being passed on to a slowly changing dimension adapter.
*The above three steps called ETL step*

**Multidimensional Data step.**
The multidimensional data model is based on the key concepts cube, dimension and hierarchy. The cube is the concept that describes the whole of the data which are presented along labeled edges of that cube, the dimensions. Whereas Hierarchies define the way in which dimensions are grouped. This dimensional modelling approach results in a database design that is consistent with the paths by which users wish to enter and navigate a cancer data warehouse. In this case, cubes, dimensions, measures, hierarchies, levels and cells constitute the basic OLAP structures. These, taken together, define the logical structure of an OLAP database. Measures are the data that we wish to analyze, while dimensions define the organization of these measures [15].

Our Cancer data warehouse may contain a FactMedical table that has fields for Patients, Time, Treatments, Procedures, Cost and Quantity. If so, we will generally analyze Cost and Quantity by warehouse, Patients, Time, Treatments and Procedures. In this case, Cost and Quantity will be our measures, and warehouse, Patients, Time, Treatments and Procedures will each be a dimension. The dimension contains the elements that called members. The major operations that could be done on OLAP cubes are Selection, Roll-up, Drill-Down and Slice, through which we can view data from all perspectives and all levels [16].

International Journal of Database Management Systems ( IJDMS ) Vol.5, No.3, June 2013

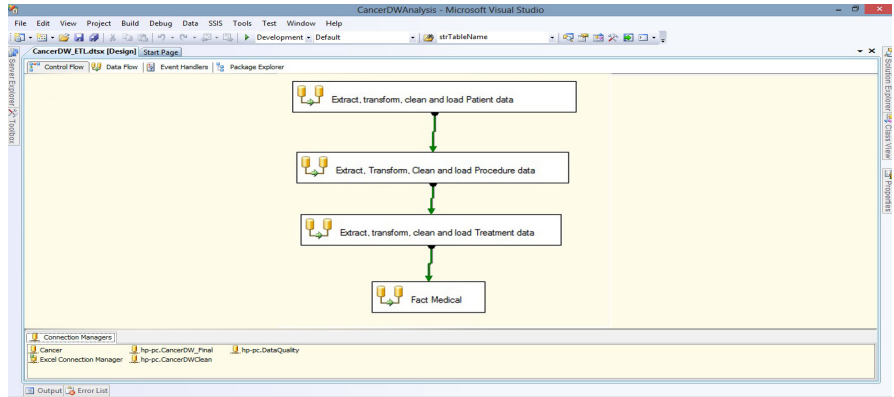

Figure 3: ETL package

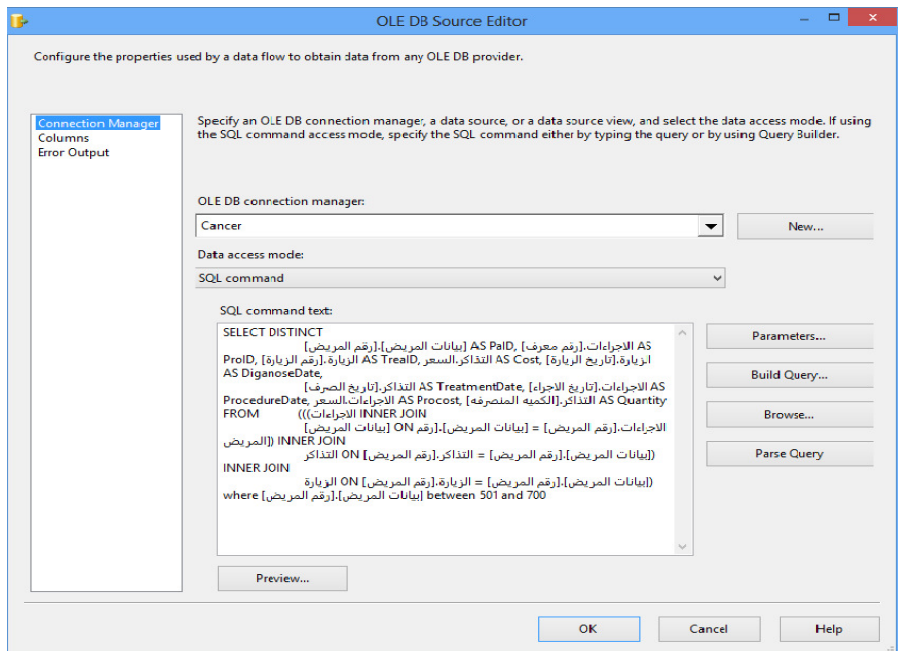

Figure 4: Extract data for FactMedical fact table

International Journal of Database Management Systems ( IJDMS ) Vol.5, No.3, June 2013

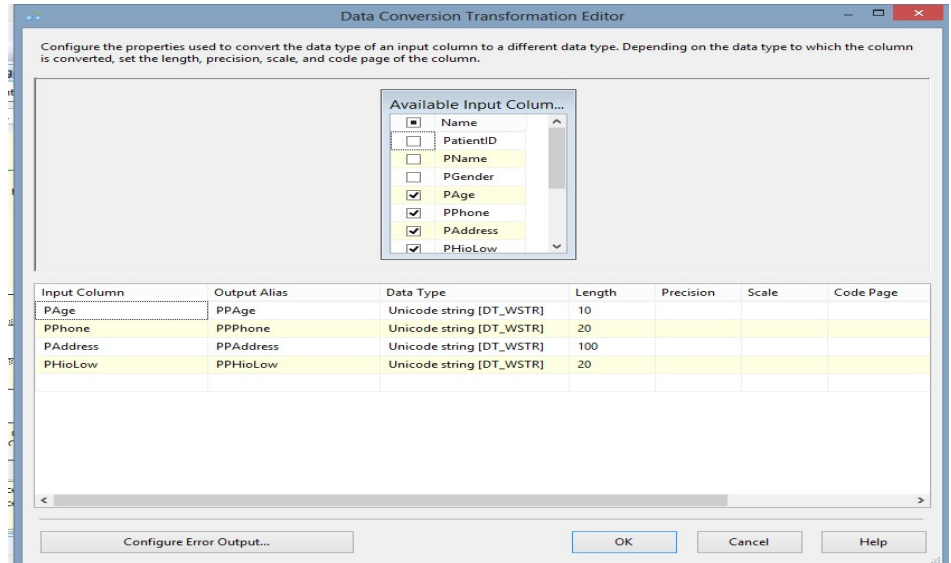

Figure 5: Transformation patient data

## 4. IMPLEMENTATION

Cancer data warehouse was used in Health Insurance Organization – Elsharkiya Branch - Egypt. The results enable the prototype properties to be validated and demonstrate the capabilities of the system in real application case. As an illustrated example, the cost of treatment for Cancer diseases is shown in the following paragraphs. The procedure for other decision-making in healthcare is similar.

**The features in our Cancer data warehouse:**

**The interface is for novices users shown in figure 6.**
The main window contains the menu bar and tools bar that call all other application forms. There we explain all elements in this main window. The first menu "Create Cancer DW" contains the submenu "Create Star Schema, Create Fact Table, Create Dimension Table and Create Relationship". The Second menu "Extract - Transform - Load" contains the submenu "Execute ETL Package". The Third menu "Multidimensional Data" contains the submenu "Create Multidimensional Data and Deploy Multidimensional Data". The fourth menu "Information Delivery" contains the submenu "Patient and Medical Procedure Information, Patient Information, Treatment Information and More Report over Web".

**Display views:** The query results are displayed in several formats, including line chart, pie chart, bar chart, area chart and report file.

**Result display:** results will be displayed on the interface. Which using C# language MDX query and Excel, is shown in Figure 7.

It facilitates significantly the managers to formulate an appropriate healthcare decision or warehouse storage strategy.

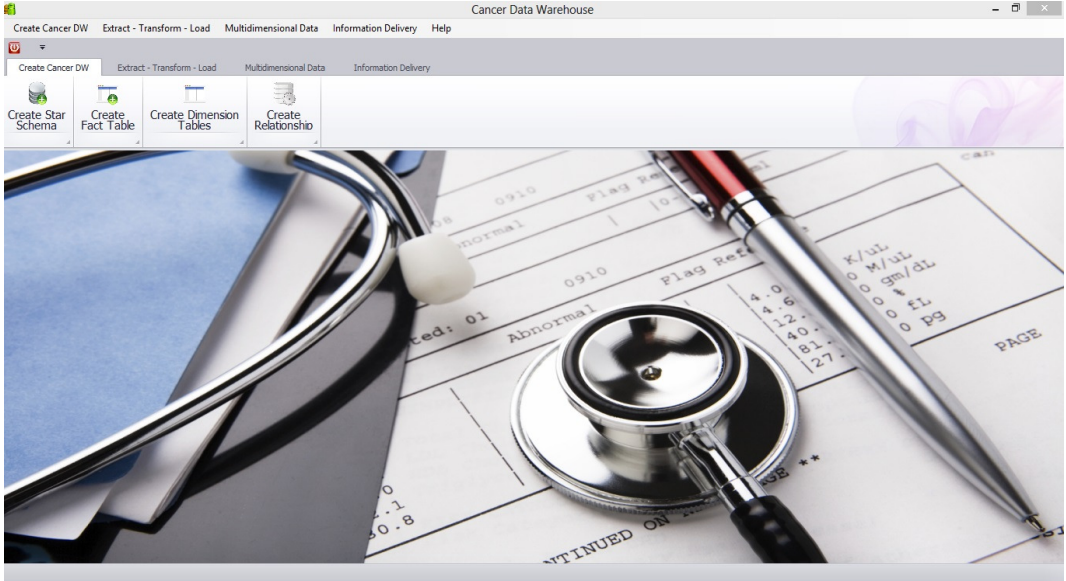

Figure 6: Cancer Data warehouse user interface

a) Show cost for each health insurance organization law (HIO Law) in pivot table and chart show number of patients by sex (green column = Male and orange column=Female) shows in C# App.

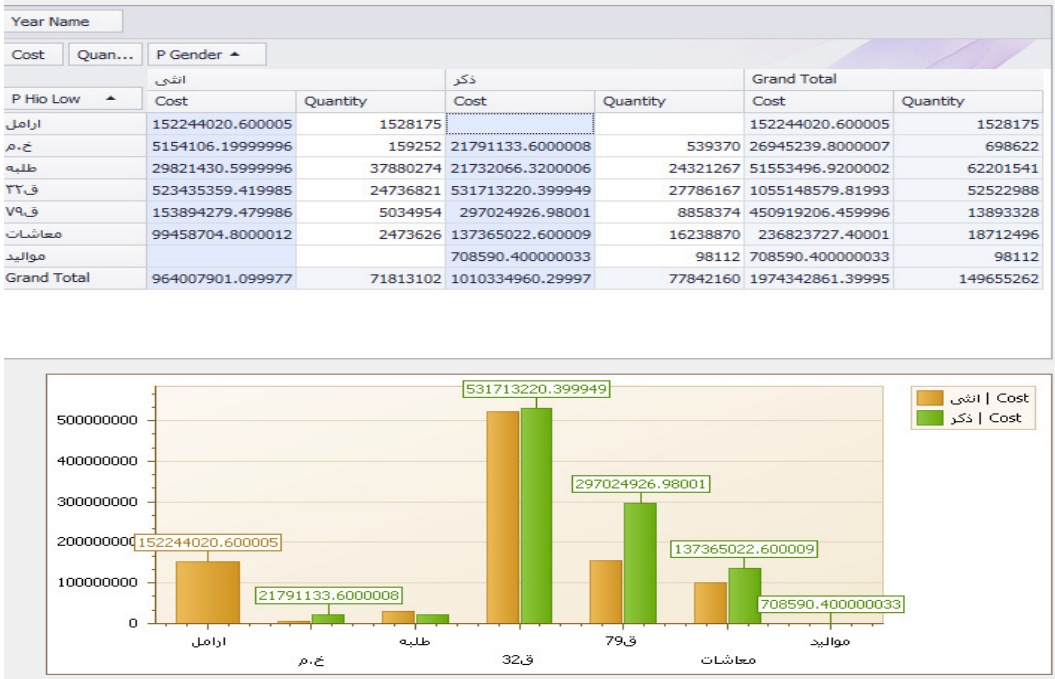

Figure 7: Cancer diseases observed from various views (a –e)

b) Cost time series shows in C# Application

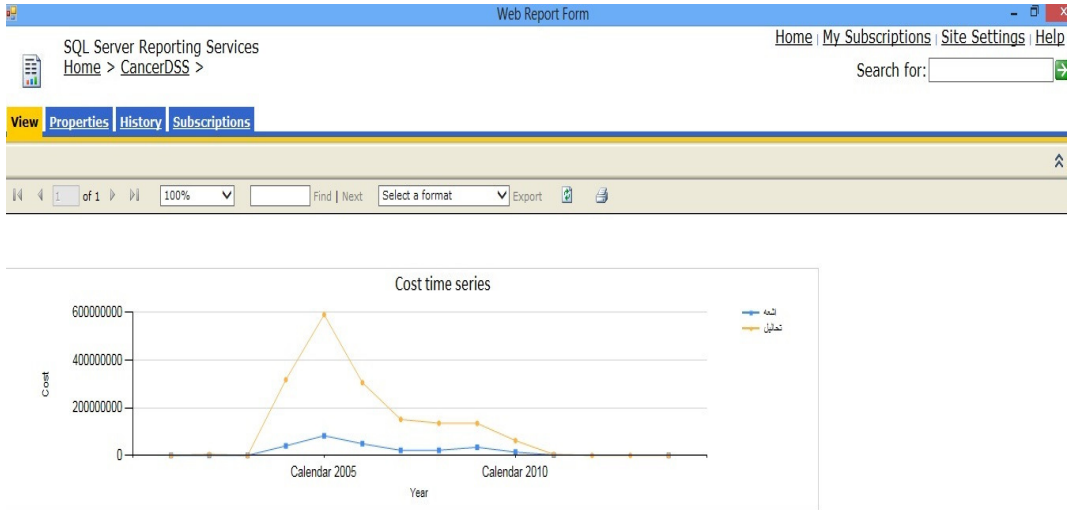

c) Right above screen show number of procedure (rays and medical tests) by year and the right blew screen show procedure time series (green line = medical tests and orange line= rays)

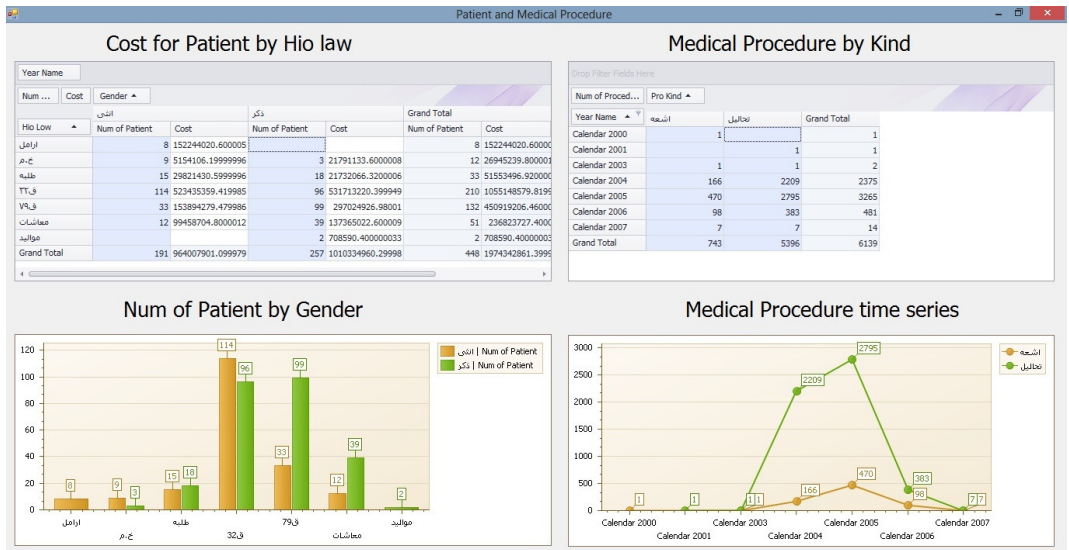

Figure 7: (continued)

d) Column chart show the number of medical procedure by patient gender

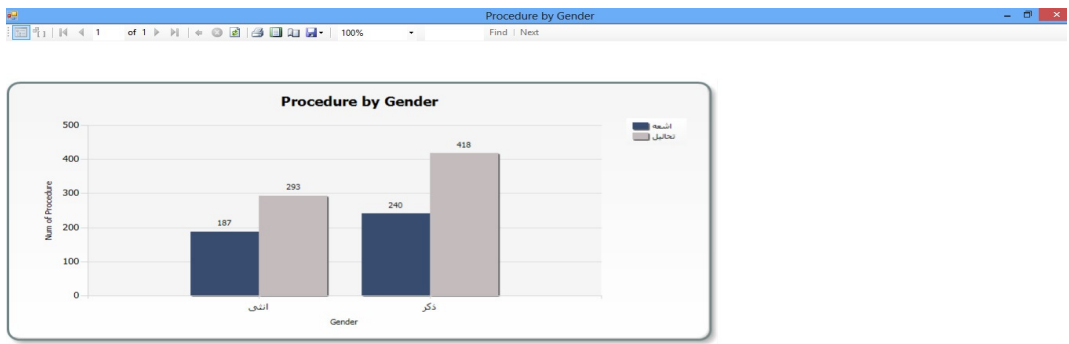

e) Pie chart in Excel show the patient cost by HIO law

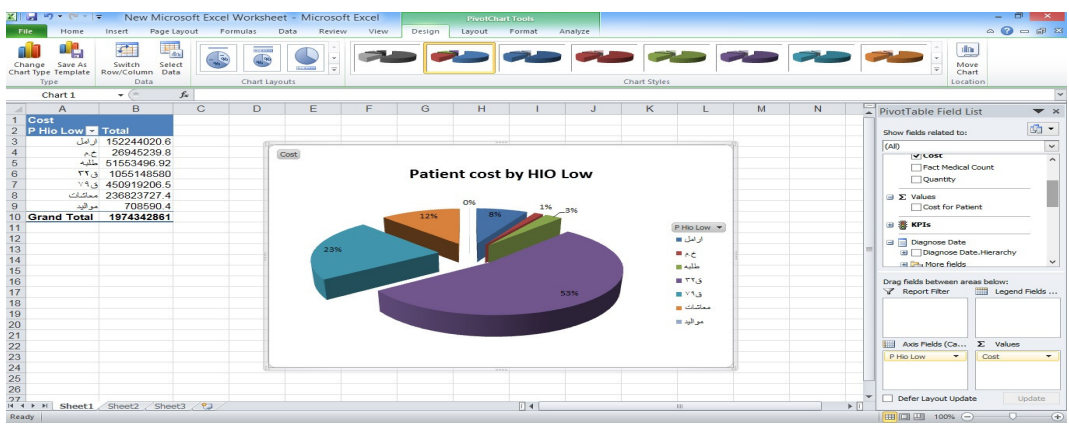

Figure 6:  (continued)

## 5. CONCLUSION

The using technology of ETL, OLAP and reporting that present in this paper to build software component. The implementation tools are made on Microsoft Sql Server 2008, SSIS, SSAS and SSRS. Cancer data warehouse is advanced at least in the following aspects: this enables insight to be gained into the factors having impacts on healthcare management activities that will help managers in making decisions to improve management performance. Also provides extremely fast response to queries. Cancer data warehouse is multidimensional. Users can view figures from multiple perspectives. In short, Cancer data warehouse is able to assist executive managers and doctors by providing accurate and timely information for Healthcare decision-making.

## AUTHORS

Dr. Osama E. Sheta; PhD (Irkutsk State Technical University, Irkutsk, Russia). Assistant Professor (IS) Faculty of Science, Zagazig University, Egypt. Assistant Professor (IS) Faculty of Science, Al Baha University, KSA.

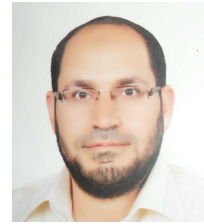

Ahmed Nour Eldeen; BS (Mathematics and Computer Science); University of Zagazig, Egypt (2006) and pre-MSc titles in computer science from Zagazig University ( 2011). Director of information systems at Health Insurance Organization Egypt (Elsharkiya branch) and has a strong background in Relational Database Systems, Database Administration , ASP.NET and C# language.

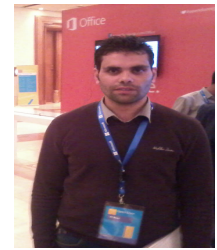